\documentclass[intlimits,twoside,a4paper]{article}
\usepackage[cp1251]{inputenc}

\usepackage[eqsecnum]{cmpj3}

\usepackage{bm}

\issue{2019}{22}{3}{33301}
\doinumber{10.5488/CMP.22.33301}

\title[The physical studies and interaction with anti-apoptotic proteins]%
{The physical studies and interaction with anti-apoptotic proteins of 2-(bis(cyanomethyl)amino)-2-oxoethyl methacrylate molecule\thanks{Corresponding authors: Serap Yalcin, syalcin@ahievran.edu.tr, Emine Babur Sas, ebsas@ahievran.edu.tr.}}
\author[S. Yalcin \textsl{et al.}]{S. Yalcin\refaddr{label1},
        E.B. Sas\refaddr{label2}, N. Cankaya\refaddr{label3}, F. Ercan\refaddr{label4}, M. Kurt\refaddr{label5} }
\addresses{
\addr{label1} K{\i}rsehir Ahi Evran University, Department of Molecular Biology and Genetic, K{\i}rsehir, Turkey
\addr{label2} K{\i}rsehir Ahi Evran University, Technical Sciences Vocational Schools,  K{\i}rsehir, Turkey
\addr{label3} Usak University, Department of Chemistry, Usak, Turkey
\addr{label4} K{\i}rsehir Ahi Evran University, Department of Plant Production, K{\i}rsehir, Turkey
\addr{label5} K{\i}rsehir Ahi Evran University, Department of Physics, K{\i}rsehir, Turkey
}

\date{Received April 17, 2019, in final form July 12, 2019}

\begin{document}

\maketitle

\begin{abstract}
In this work 2-(bis(cyanomethyl)amino)-2-oxoethyl methacrylate (CMA2OEM) molecule has been characterized theoretically. First, the potential energy surface has been calculated to find the lowest energy state of the molecule. After the most stable state of the molecule, Mulliken atomic charge and nonlinear-optical properties were investigated. Also in the study, binding poses of CMA2OEM molecule and anti-apoptotic proteins, such as BCL-2, BCL-w, MCL-1, AKT1 and BRAF were investigated. The molecular docking results showed that the most stable complex was obtained with this molecule and BRAF protein. The molecular docking results showed that the most stable complex was obtained with this molecule and serine/threonine-protein kinase protein. This study suggested that molecular docking approach may be a potential tool to identify the hydrogen bond interactions in order to treat a disease. Finally, this new ligand could pave the way to experimental studies.

\pacs 33.15.-e, 31.15E-
\end{abstract}

\section{Introduction}

Acrylate derivatives which are soluble in many organic solvents are widely used in bath tubs in glass materials \cite{Zub74}. The reason for this is the versatility of acrylic monomers, which indicates their adaptability. In recent studies, this turned out to be urgent for amino methacrylate derivatives in areas such as waste water treatment, biochemical sensor and protein purification \cite{Bus89}. Due to their physical properties, acrylates are used in medicine, orthopedics, tooth and filling materials, drug delivery systems, biochemical sensors and soft tissue studies \cite{Mel00}. A number of studies have been performed in our team on the synthesis of meth/acrylate monomer and their polymerization. 2-(bis(cyanomethyl)amino)-2-oxoethyl methacrylate (CMA2OEM) is one of the important monomers that can be obtained in 2 steps \cite{Mry94}.

Apoptosis is an effective mechanism for eliminating damaged or unnecessary cells by multicellular organisms \cite{Sta11}. Anti-apoptotic proteins are overexpressed in most cancers and, due to this feature, are highly remarkable as the target of anti-cancer agents \cite{Sta04}. Therefore, the study has been performed to determine the anticancer potential of 2-(bis(cyanomethyl)amino)-2-oxoethyl methacrylate (CMA2OEM) targeting anti-apoptotic proteins, thereby inducing apoptosis in cancer cells. 

In this study, physical characterization of 2-(bis(cyanomethyl)amino)-2-oxoethyl methacrylate \linebreak(CMA2OEM) compound was carried out by Sas et al. \cite{Mry94}, in which synthesis and some theoretical studies were carried out. Firstly, the potential energy surface is calculated to find the most optimized state of the molecule. After the molecular optimization is performed by the DFT-B3LYP method, the Mulliken charges, nonlinear optical properties and dipole moment are presented. We demonstrated  suitable confirmations of the new synthesis CMA2OEM compound as a ligand within anti-apoptotic proteins binding sites \textit{in silico} by using Autodock vina114 \cite{Scient2012}. \textit{In silico} studies further provided predictive binding properties of selected ligands for inhibition of target protein.

\section{Materials and methods}
\subsection{Docking procedure}
CMA2OEM compound was docked into active sites of anti-apoptotic proteins, BCL-2, BCL-w, MCL-1, AKT1 and BRAF in Autodock Vina software \cite{Scient2012}. The structure of anti-apoptotic proteins are freely available from the RCSB Protein Data Bank as a 3D theoretical model (PDB ID; BCL-2: 4man, BCL-w: 2y6w, mcl-1: 5fdo, AKT1: 4gv1, BRAF: 5vam). 2D structure of the ligand was converted to energy minimized 3D-structure. All proteins and ligand were validated before performing the \textit{in silico} computations. Interaction of amino acid residues of proteins with ligand was analyzed using LIGPLUS tool \cite{Abr64,Olver64}.

\subsection{Computational procedure}
The potential energy surface was calculated using the DFT/B3LYP/6-311++G(d,p) method to determine the most stable state of the CMA2OEM molecule \cite{Yuk87,Shv04}. To study the charge distribution of the molecule, Mulliken charges calculated using the same method are shown on the graph (figure~\ref{fig3}). Molecular polarizability, hyper polarizability and dipole moment values are calculated in the output file CMA2OEM molecule and tabulated. Gaussian 09 package program was used in these calculations \cite{url2}.

\section{Results and discussion}
\subsection{Potential energy surface (PES) scan and optimized structure }
\begin{figure}[!t]
	\centerline{\includegraphics[width=0.65\textwidth]{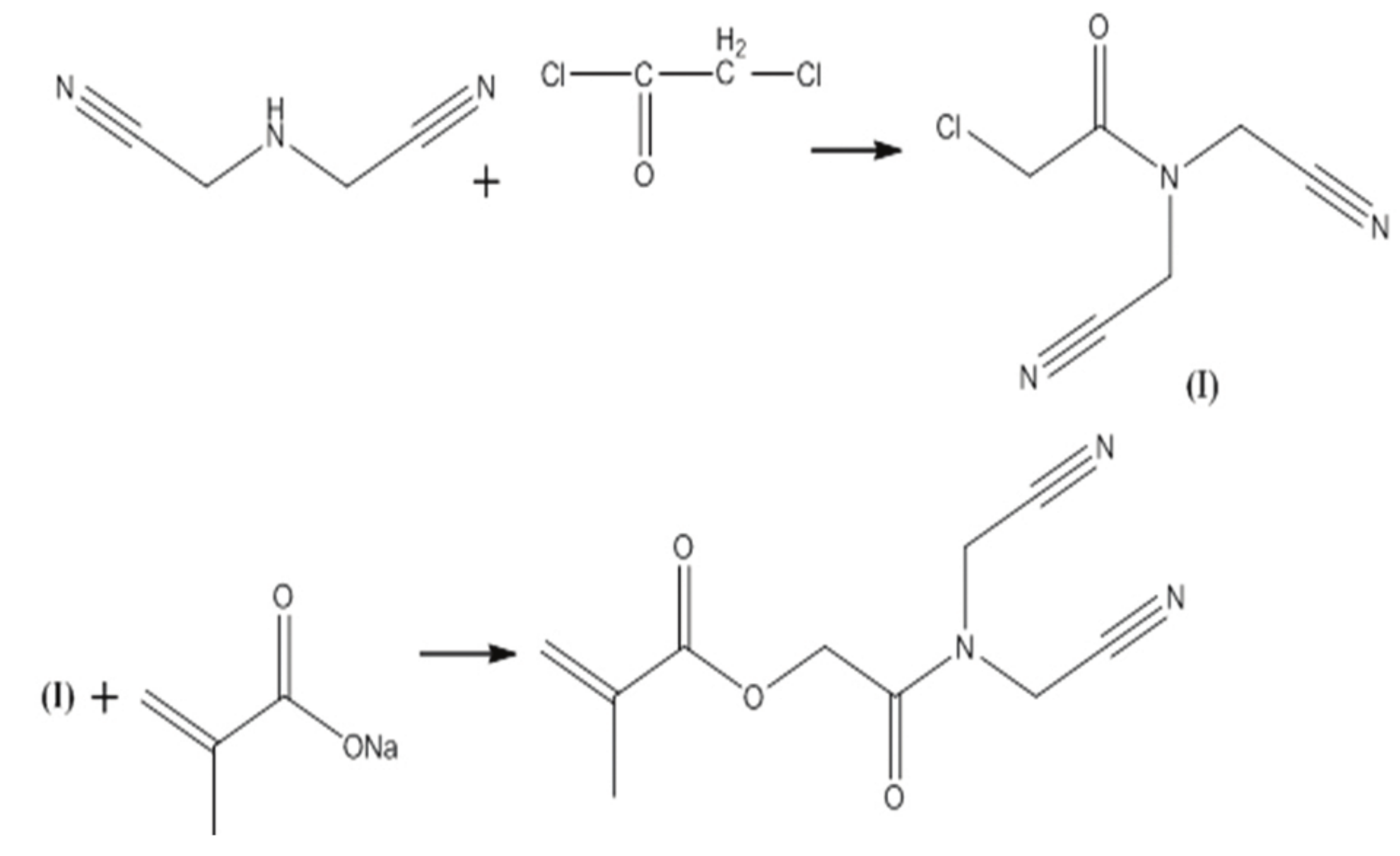}}
	\caption{Synthesis of the 2-(bis(cyanomethyl)amino)-2-oxoethyl methacrylate (CMA2OEM).} \label{fig1}
\end{figure}

\begin{figure}[!t]
 	\centerline{\includegraphics[width=0.65\textwidth]{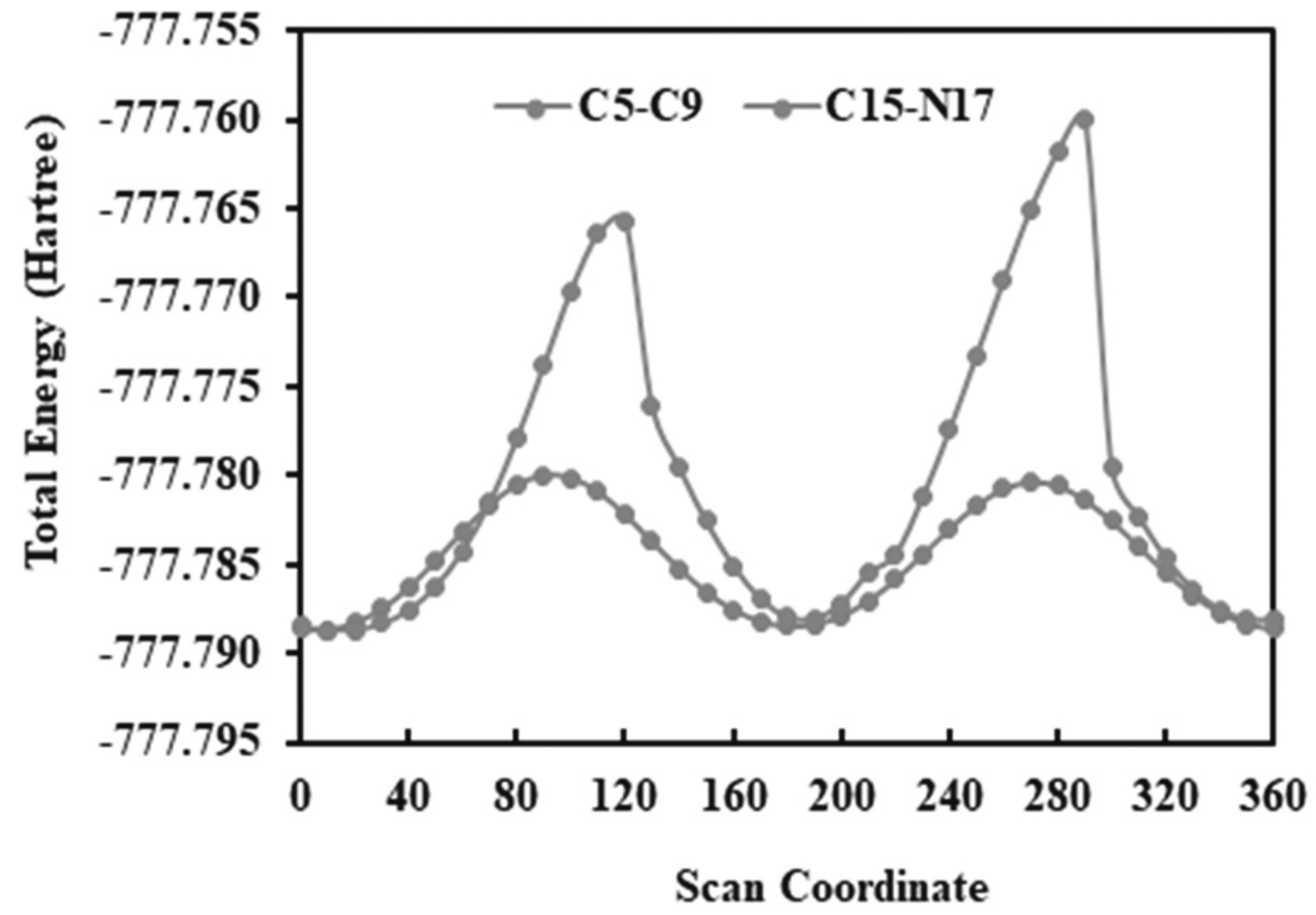}}
 	\caption{PES scan for dihedral angles O11-C9-C5-C1 and C12-C15-N17-C18 of CMA2OEM.} \label{fig2}
 \end{figure}

The PES (potential energy surface) is calculated around the single bonds so that the optimal structure of a molecule can be found. Firstly, to find the lowest energy state of  the title molecule, torsional angles were calculated between C and N atoms. Rotation around C15-N17 bond (C12-C15-N17-C21). On this calculation, torsion angle was varied from 0 degrees to 360 degrees by changing every 10 degrees, 36~steps were taken. A second torsion angle was calculated after the lowest energy state of the molecule was found. Rotation around C9-C5 bond (O11-C9-C5-C1). On this calculation, torsion angle was also varied from 0 degrees to 360 degrees by changing every 10 degrees, 36 steps were taken. Synthesis reaction of CMA2OEM \cite{Mry94} and its potential energy surfaces (PES) were shown in figures~\ref{fig1} and \ref{fig2}, respectively.

 The molecular structure of CMA2OEM in the ground state was optimized with B3LYP method with 6-311++G(d,p) basis set using the Gaussian 09 software after finding the lowest energy state \cite{Yuk87,Shv04,url2}.

\subsection{Mulliken atomic charges }
\begin{figure}[!t]
	\centerline{\includegraphics[width=0.65\textwidth]{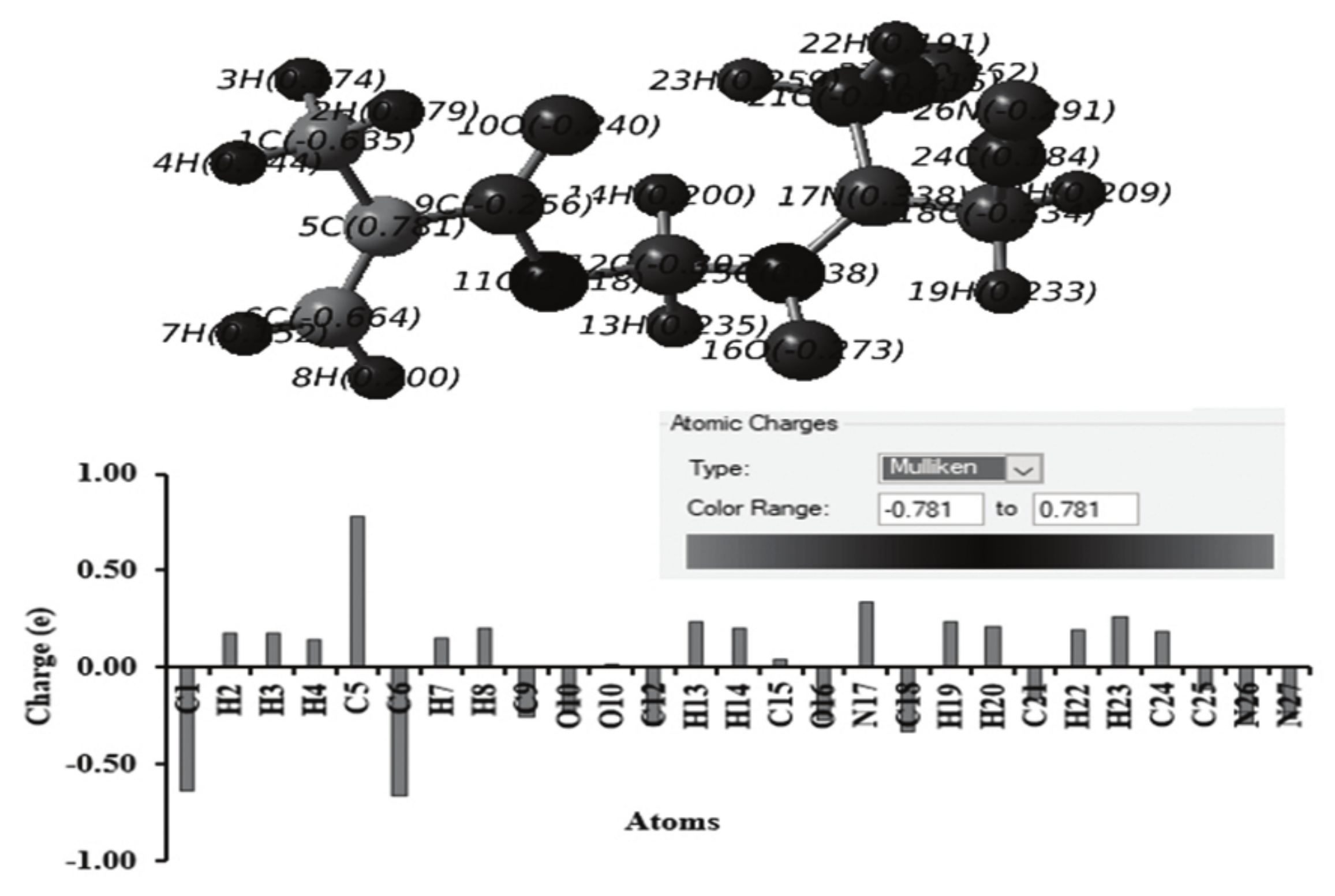}}
	\caption{The Mulliken charge distribution for CMA2OEM.} \label{fig3}
\end{figure}

The Mulliken charge distributions of CMA2OEM were calculated with B3LYP/6-311++G(d,p) method and shown in figure~\ref{fig3}.

 Mulliken charges provide physical information for the electronic distribution of the molecule. Among the carbon atoms in the molecule, C5 ($0.78e$) has the highest positive value, C1 and C6 ($-0.64e$ and $-0.66e$) have the lowest negative value. For this reason, we can say that carbon atoms of C5 and C6 regulate the molecular distribution of the charge. The oxygen and nitrogen atoms in the molecule are linked by carbon atoms. As a result, when the carbon atoms take on a variable value ($0.18e$, $-0.29e$, $-0.26e$, etc.), the nitrogen and oxygen atoms take on a value of about $-0.2e$, except for the oxygen atom between the two carbon atoms (C9 and C12). This distribution may be due to the fact that these atoms are less electronegative than nitrogen and oxygen because they concentrate on carbon atoms.

\subsection{Nonlinear optical properties and dipole moment }
In organic materials, optical properties are determined by polarizability. The polarizability of an atom or molecule is a measure of how easily the nucleus and electrons can displace their stable state. It is the valence electrons farthest from the nucleus of electrons that are easily displaced in an atom or molecule. For this reason, the valence electrons have a great contribution to the polarizability. The bonds between carbon atoms and other elements are of two kinds, $\sigma$ and $\pi$ bond. The nonlinear optical properties of molecular systems depend on the polarizability of electrons in the $\pi$-bond. Polarizability discloses the electronic structure of the molecule precisely and comprehensibly.  Nonlinear optical properties (NLO) viz., molecular polarizability ($\alpha$), anisotropy of polarizability ($\Delta\alpha$), molecular first hyperpolarizability~($\beta$) and electronic dipole moment  ($\mu$) for the study compound were evaluated. The polarizability ($\alpha$), anisotropy of polarizability ($\Delta\alpha$), molecular first hyperpolarizability ($\beta$) are obtained by the following equation and the values are listed in table~\ref{tab1} \cite{han12,14,15}.
\begin{align}
\alpha&= \frac{1}{3}(\alpha_{xx}  + \alpha_{yy} +\alpha_{zz}), \nonumber\\
\Delta \alpha&= \frac{1}{2}\big[ (\alpha_{xx} - \alpha_{yy})^{2} + (\alpha_{xx} - \alpha_{zz})^{2} + (\alpha_{yy} - \alpha_{zz})^{2}\big],  \nonumber\\
\beta &= (\beta_{xxx} + \beta_{xyy} + \beta_{xzz})^{2}  + (\beta_{yyy} + \beta_{yzz} + \beta_{yxx})^{2} + (\beta_{zzz} + \beta_{zxx} + \beta_{zyy})^{2}. \nonumber 
\end{align}

\begin{table}[!t]
	\caption{Dipole moments $\mu$ (D), polarizability $\alpha$,  anisotropy of  polarizability $\Delta\alpha$, and  first hyperpolarizability $\beta$ of CMA2OEM.}
	\label{tab1}
	\vspace{2ex}
	\begin{center}
		\begin{tabular}{c c c c}
			\hline
			\hline
			$\mu_{x}$&$-2.9967$&$\beta_{xxx}$&$2201.5729$\strut\\
			$\mu_{y}$&$-1.3147$&$\beta_{xxy}$&$-469.5798$\strut\\
			$\mu_{z}$&$0.7051$&$\beta_{xyy}$&$688.1708$\strut\\
			$\mu_{0}$&$3.347508475$&$\beta_{yyy}$&$-342.3852$\strut\\
			$\alpha_{xx}$&$23.510075$&$\beta_{xxz}$&$329.6725$\strut\\
			$\alpha_{xy}$&$-1.437457$&$\beta_{xyz}$&$90.3300$\strut\\
			$\alpha_{yy}$&$19.352970$&$\beta_{yyz}$&$-530.3426$\strut\\
			$\alpha_{xz}$&$-2.205644$&$\beta_{xzz}$&$-489.2093$\strut\\
			$\alpha_{yz}$&$-0.756278$&$\beta_{yzz}$&$-177.4980$\strut\\
			$\alpha_{zz}$&$19.913798$&$\beta_{zzz}$&$262.9360$\strut\\
			$\alpha$&$20.925614$&$\beta_{x}$&$2400.53445$\strut\\
			$\Delta \alpha$&$40.90764548$&$\beta_{y}$&$-2.193229717$\strut\\
			&&$\beta_{z}$&$62.26591199$\strut\\
			&&$\beta$&$2597.206142$\strut\\
			\hline
		\hline
		\end{tabular}
	\end{center}
\end{table}

The calculated parameters and electronic dipole moment for CMA2OEM are tabulated in table~\ref{tab1}. It is well known that the higher values of dipole moment, molecular polarizability, and hyperpolarizability are important for more active NLO properties. CMA2OEM has a relatively homogeneous charge distribution and it does not have a large dipole moment. The calculated value of dipole moment was found to be 3.3475~D. If we compare the common values of urea ($\alpha$ and $\beta$ of urea are 1.3732 D and 	$372.89\cdot 10^{-33}$~esu) the hyperpolarizability and dipole moment values of CMA2OEM are larger than those of urea.

The conversion coefficients are as follows:
polarizability $\alpha=1~\text{au} = 0.1482\cdot10^{-24}$~esu,
first hyperpolarizability $\beta=1~\text{au} =  0.863993\cdot10^{-32} $~esu.

\subsection{Molecular docking }
Molecular docking calculations were obtained from two different programs: Autodock Vina and VMD \cite{16}. Water molecules and cofactors were removed from the proteins to provide an interaction between only ligand and receptor. The Lamarckian Generic Algorithm was used as a score function to guess the best interaction between ligand and anti-apoptotic proteins. The highest binding score refers to the most stringent binding between protein and ligand. The docking results calculated by Vina  were presented in table~\ref{tab2}. According to these results, the highest binding score was obtained between CMA2OEM molecule and anti-apoptotic BRAF protein with $-6.5$~kcal/mol affinity energy.

\begin{table}[!b]
	\caption{Docking binding energy results of novel CMA2OEM molecule as inhibitor with anti-apoptotic proteins.}
	\label{tab2}
\vspace{2ex}
\begin{center}
	\begin{tabular}{c c }
		\hline\hline
	CMA2OEM&Affinity energy	(kcal/mol)
	\strut\\
		\hline
		\hline
		BRAF (PDB ID: 5vam)&$-6.5$\strut\\
		BCL-2 (PDB ID: 4man)&$-5.9$\strut\\
		BCL-w (PDB ID: 2y6w)&$-6.1$\strut\\
		mcl-1 (PDB ID: 5fdo)&$-6.4$\strut\\
		AKT1 (PDB ID: 4gv1)&$-6.3$\strut\\
		\hline
		\hline
	\end{tabular}
\end{center}
\end{table}

\begin{figure}[!b]
	\centerline{\includegraphics[width=0.64\textwidth]{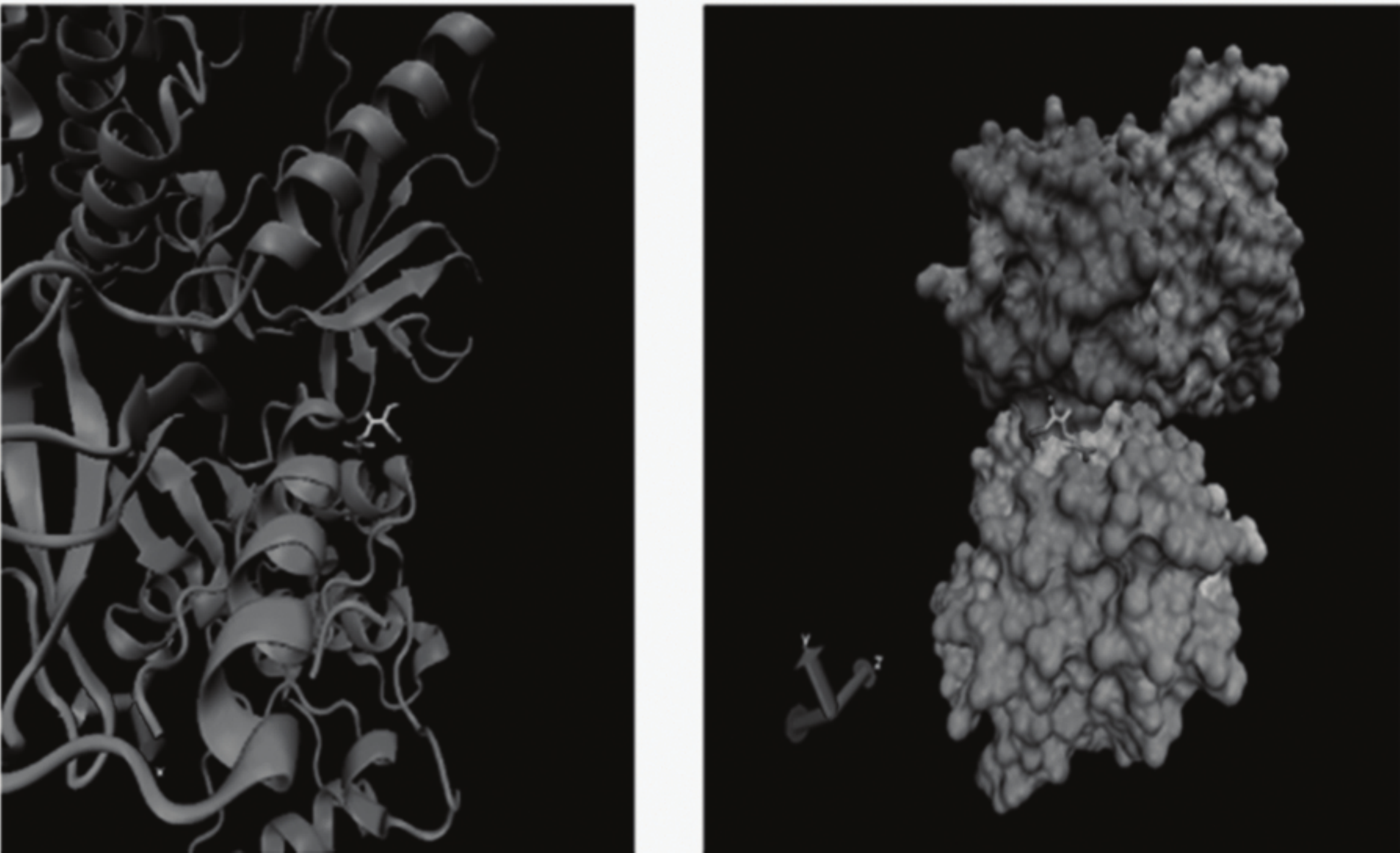}}
	\caption{Docking figure of CMA2OEM molecule in BRAF protein cavity. } \label{fig4}
\end{figure}

CMA2OEM was found to be docked at all tested anti-apoptotic proteins with good confirmation. BRAF and CMA2OEM interactions can be observed in figure~\ref{fig4}. In the BRAF and CMA2OEM compound (table~\ref{tab3}), hydrogen bonds with oxygen atom of the ester and THR 458, GLN 562 and GLY 478 residues were identified. Van der Waals interactions with amino acid residues GLU 715, ALA 712, ASP 565, LEU 711, HIS 477, LYS 475 and TRP 476 were also determined.

\begin{table}[!t]
	\caption{Interacting amino acid residues of anti-apoptotic proteins with ligand \cite{17}.}
	\label{tab3}
	\begin{center}
		\begin{tabular}{c c }
			\hline\hline
			Number of hydrogen bonds&Types of amino acid-ligand interaction 	\strut\\
			\hline
			\hline
			3&GLN 562, GLY 478, THR 458\strut\\
			1&ARG 124\strut\\
			1&ARG 95\strut\\
			1&THR 195\strut\\
			\hline
			\hline
		\end{tabular}
	\end{center}
\end{table}

 \begin{figure}[!t]
	\centerline{\includegraphics[width=0.62\textwidth]{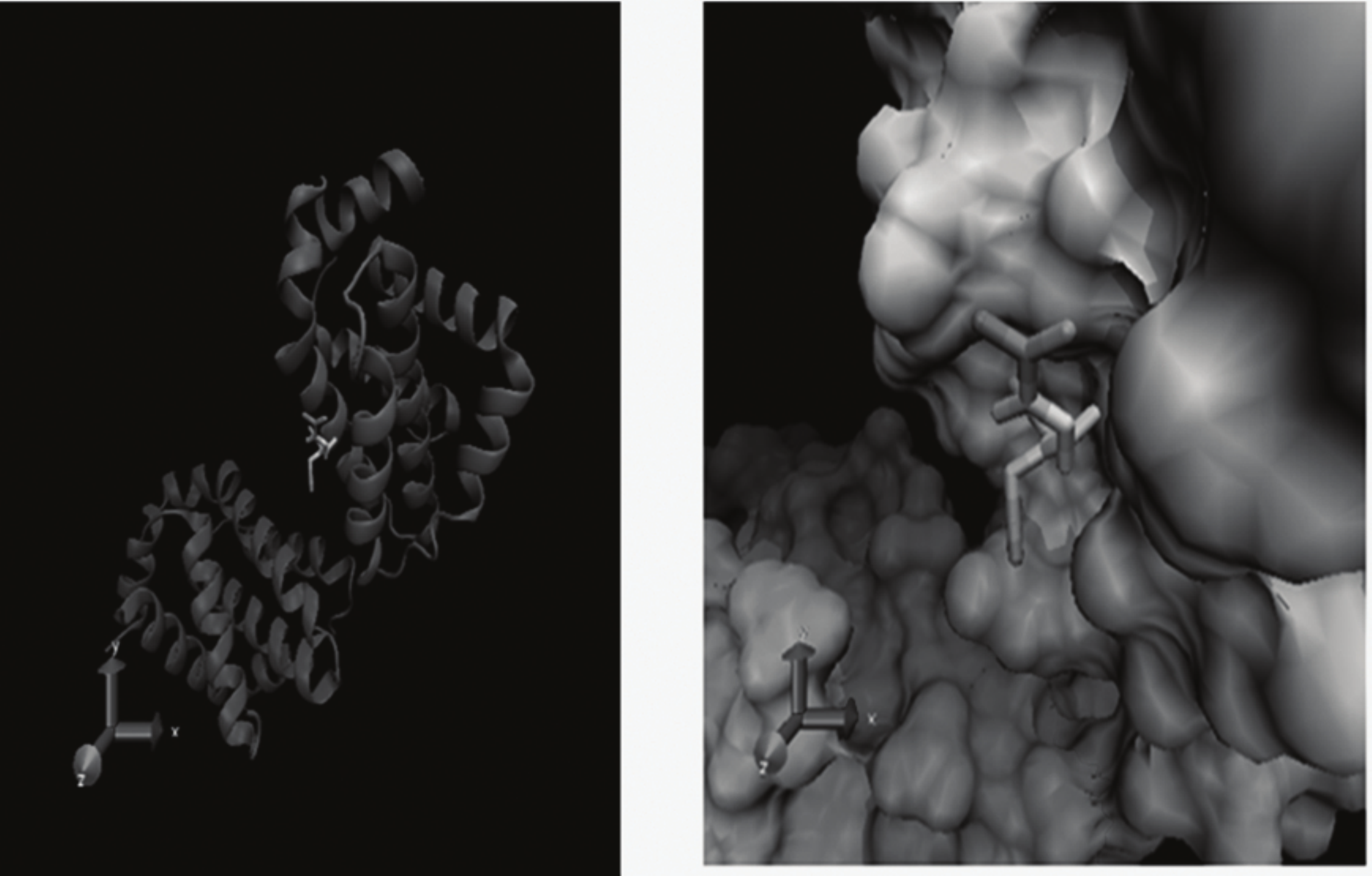}}
	\caption{Docking figure of CMA2OEM molecule in Bcl-2 protein cavity.} \label{fig5}
\end{figure}

\begin{figure}[!t]
	\centerline{\includegraphics[width=0.62\textwidth]{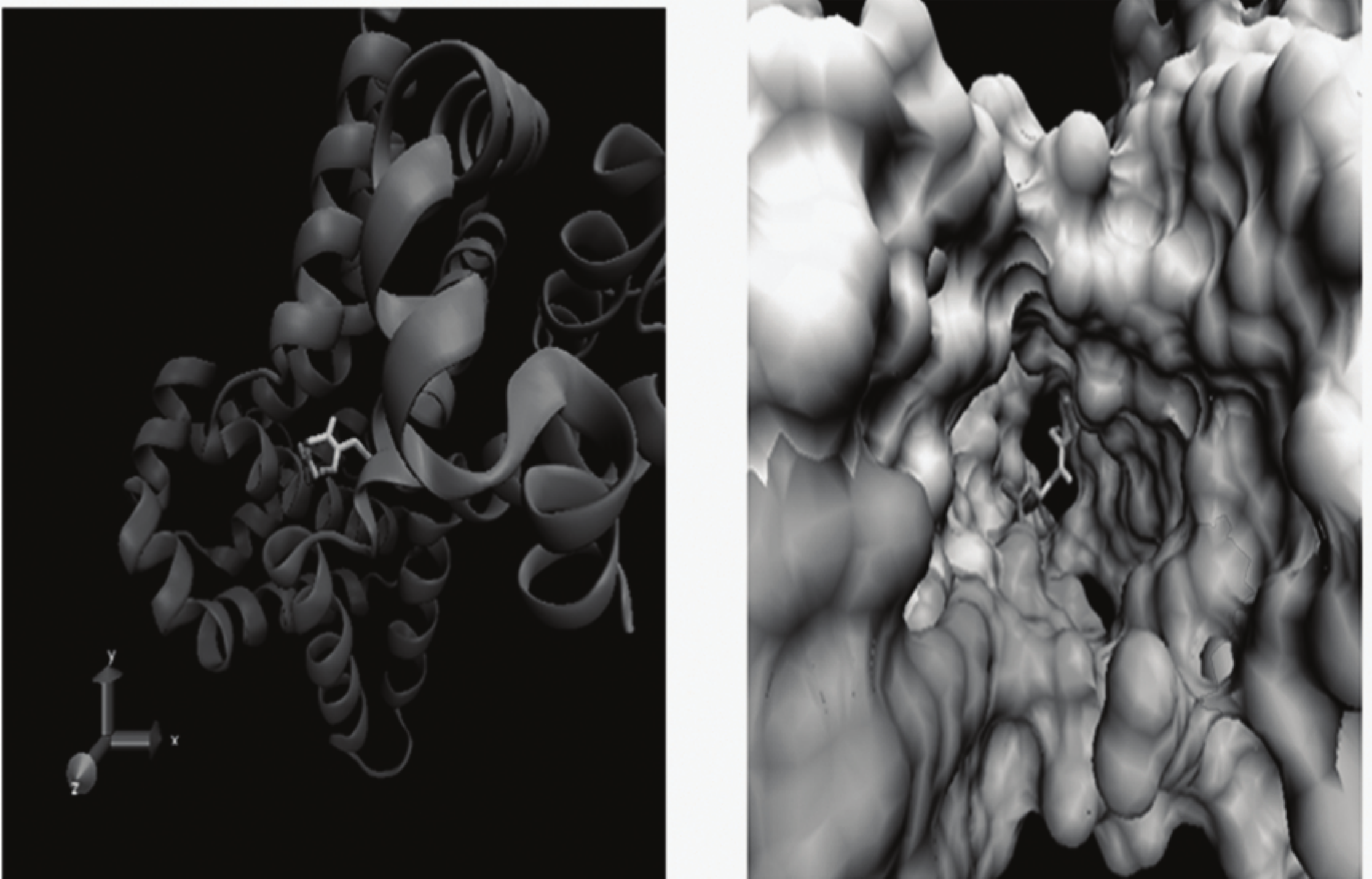}}
	\caption{Docking figure of CMA2OEM molecule in Bcl-w protein cavity.} \label{fig6}
\end{figure}

\begin{figure}[!t]
	\centerline{\includegraphics[width=0.62\textwidth]{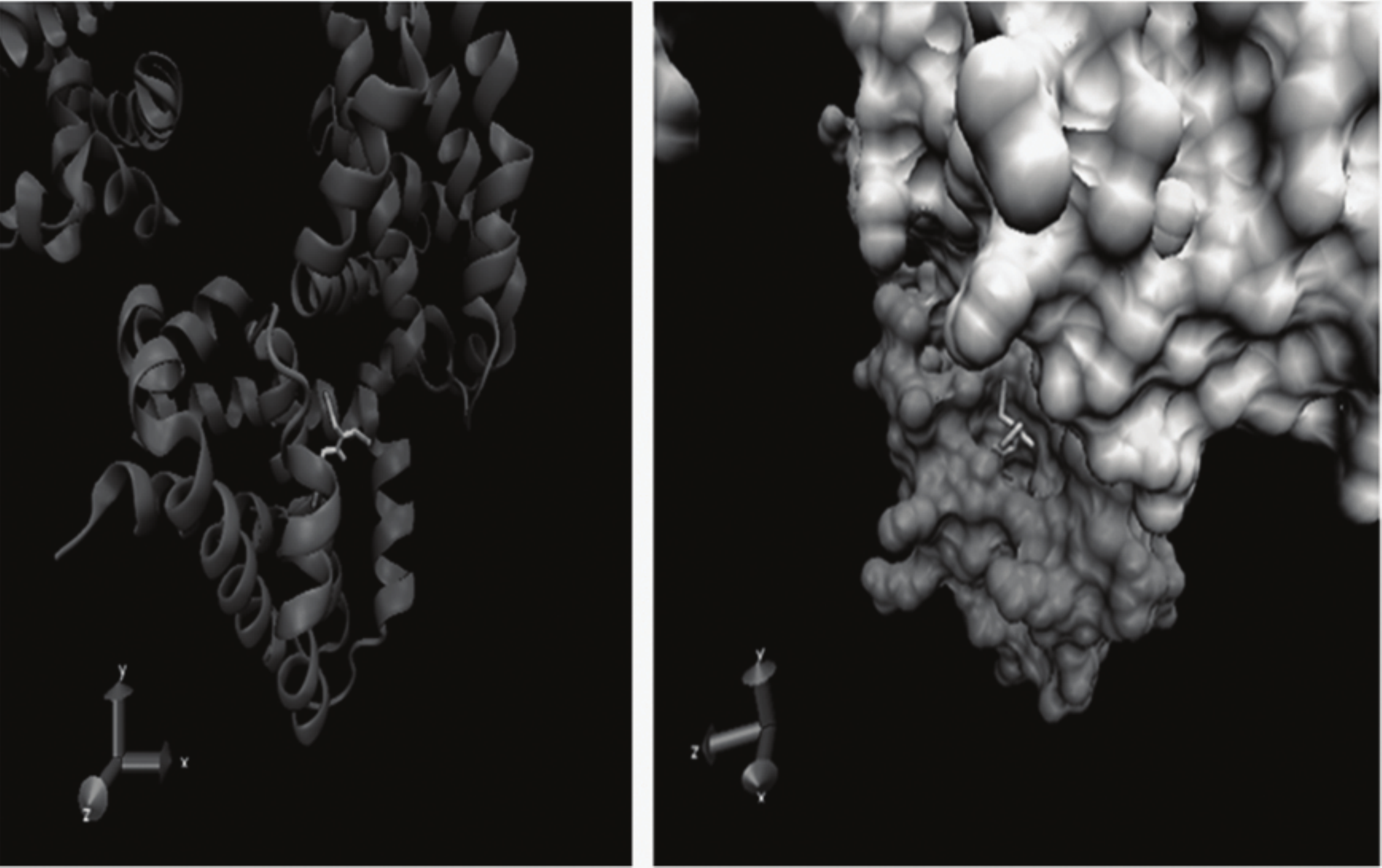}}
	\caption{Docking figure of CMA2OEM molecule in Mcl-1 protein cavity.} \label{fig7}
\end{figure}

\begin{figure}[!t]
	\centerline{\includegraphics[width=0.62\textwidth]{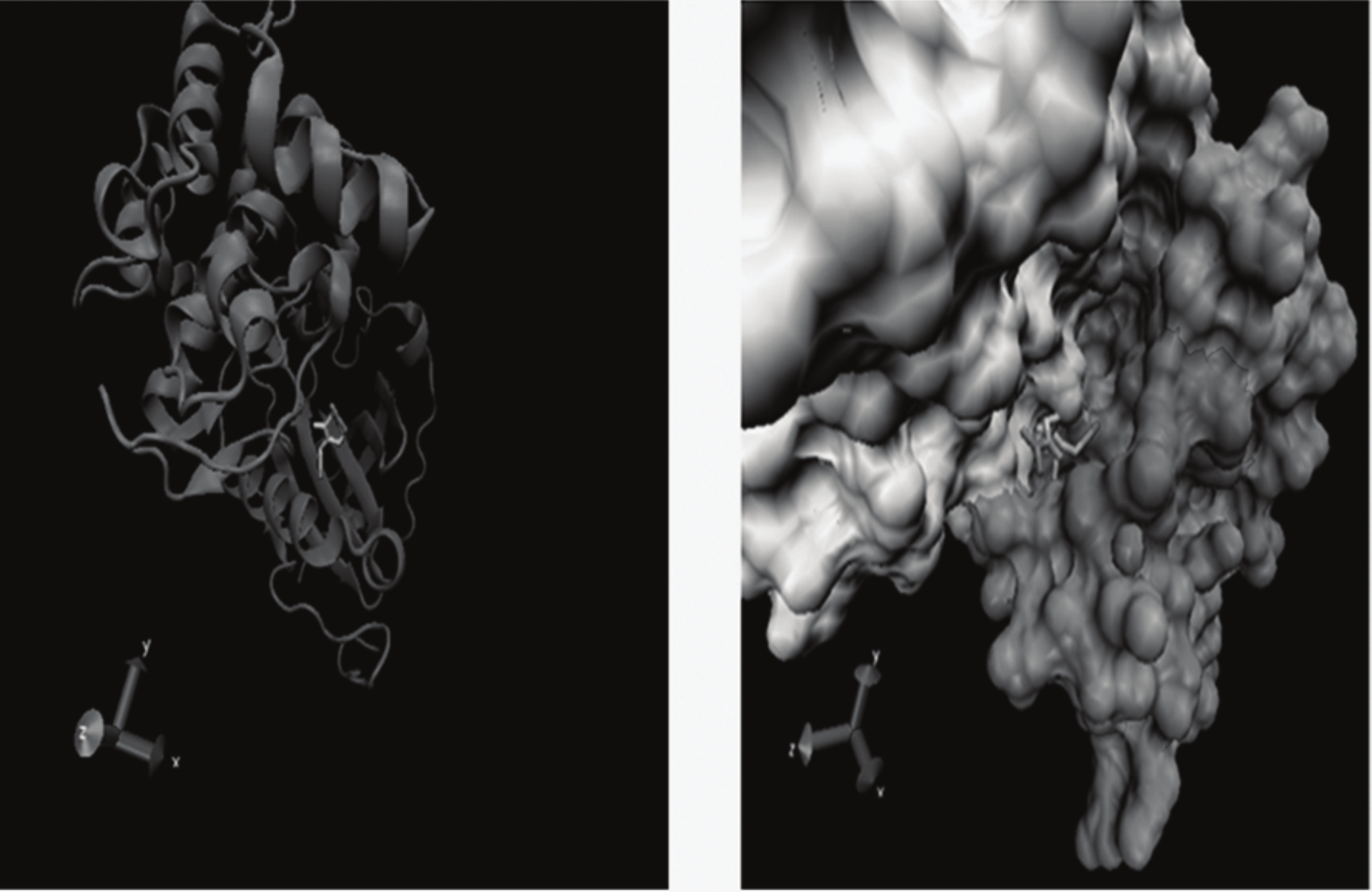}}
	\caption{Docking figure of CMA2OEM molecule in AKT1 protein cavity. } \label{fig8}
\end{figure}

Bcl-2 and ligand interaction can be observed in figure~\ref{fig5}. CMA2OEM molecule formed one hydrogen bond with ARG 124 amino acid residue and seven Van der Waals interactions with different amino acid residues. In Bcl-w and CMA2OEM interaction, hydrogen bond occurred in the same amino acid with the different residue (ARG 95) (figure~~\ref{fig6}, table~\ref{tab3}).

Mcl-1 and CMA2OEM interactions can be observed in figure~\ref{fig7}. In the Mcl-1 and CMA2OEM compound, hydrogen bond with oxygen atom of the ester and LEU 267 residue was identified. Van der Waals interactions with amino acid residues MET 250, PHE 254, VAL 253, THR 266, ARG 263, PHE 228 and PHE 270 were also determined. AKT1 and CMA2OEM interaction is shown in figure~\ref{fig8}. For AKT 1 protein and CMA2OEM, an interaction of hydrogen bond with THR 195 was identified (table~\ref{tab3}). Van der Waals interactions with amino acid residues PHE 161, GLY 294, GLU 191, HIS 194, LEU 181, LYS 179, ASP 292 and LEU 295 were also defined.

\section{Conclusion}
In this study, biological and physical characterization of CMA2OEM compound was carried out using DFT/B3LYP/6-311++G(d,p) method. After optimization of the molecule, Mulliken atomic charge and NLO properties were examined. Mulliken charges (charges distribution-chk file) graphics were drawn to  understand the properties and the dynamics of the molecule. The NLO properties of the molecule were calculated by the same method and 2 CMA2OEM was seen to be more polarized compared to the polarized molecule urea.

The possible docking alternatives were studied in Autodock Vina between CMA2OEM, as an inhibitor, and human anti-apoptotic proteins, BCL-2, BCL-w, MCL-1, AKT1 and BRAF. Inhibition capability of this molecule on these proteins was evaluated, and potential inhibition of CMA2OEM molecule was tested \textit{in silico}. The most promising result was obtained from CMA2OEM and BRAF interactions. In this interaction, more stable conformation with lower energy in ligand-protein complex was analyzed. Hence, binding studies have been shown to be a useful tool that reveals electronic affinity and can help to understand ligand-protein interactions. 

In the present study, we have designed and analysed a ligand in order to obtain a new drug active molecule. As concerns the previous literature precedence, the  aforementioned investigation has not been reported in the literature so far. Thus, this study shows that the molecular interaction affinities between anti-apoptotic targets and compound are based on molecular docking. In aggregate, these molecular docking results will aid in better understanding of its molecular action with anti-apoptotic proteins. Taken together, our findings shed light on the molecular basis of the factors governing the binding of anti-apoptotic proteins and  causes major consequences for the development of efficient therapeutic approaches.

\vspace{-3mm}
\section*{Acknowledgements}

This work was supported by Ahi Evran University Scientific Project Unit (BAP) with, Project No:~PYO–FEN.4001.15.012.

\ukrainianpart

\title{Фізичні дослідження і взаємозв'язок з антиапоптотичними білками молекули 2-(бі(ціаометил)аміно)-2-оксоетил метакрилат }
\author{С. Ялджин\refaddr{label1},  E.Б. Сас\refaddr{label2}, Н. Джанкайа\refaddr{label3}, Ф. Ерджан\refaddr{label4}, M. Курт\refaddr{label5} }
\addresses{
\addr{label1} Університет  м. Кіршехір, відділ молекулярної біології та генетики,  Кіршехір, Туреччина
\addr{label2} Університет  м. Кіршехір, професійно-технічний коледж, Кіршехір, Туреччина
\addr{label3} Університет м. Усак, відділ хімії, Усак, Туреччина
\addr{label4} Університет  м. Кіршехір,  відділ рослинництва, Кіршехір, Туреччина
\addr{label5} Університет  м. Кіршехір, відділ фізики, Кіршехір, Туреччина
}

\makeukrtitle

\begin{abstract}
У цій роботі  молекула 	2-(бі(ціаометил)аміно)-2-оксоетил метакрилат (CMA2OEM) вивчалася теоретично. Спершу розраховувалася потенціальна енергія поверхні, щоб знайти стан молекули з найнижчою енергією. Далі досліджувався найстійкіший стан молекули, атомний заряд Міллікена    і нелінійні оптичні властивості.  Також вивчалися положення зв'язків молекули CMA2OEM і антиапоптотичних білків, таких як  BCL-2, BCL-w, MCL-1, AKT1 і  BRAF. Результати молекулярного докінгу показали, що найстійкіший комплекс отримується з цієї молекули і   білка серин/треонін-протеїнкіназа. Дане дослідження дозволяє зробити припущення, що метод молекулярного докінгу  може бути потенційним інструментом для виявлення взаємодій водневих зв’язків з метою  лікування захворювання. Нарешті, цей новий ліганд міг би прокласти шлях до експериментальних досліджень.

\keywords атомний заряд Міллікена, нелінійні оптичні властивості, молекулярний докінг, антиапоптотичний, CMA2OEM

\end{abstract}

\end{document}